 \documentclass[prd,aps,preprint,superscriptaddress,showkeys,showpacs,
 preprintnumbers,amsmath,amssymb,floatfix]{revtex4-1}
\usepackage{graphicx}
\usepackage{epsfig}
\usepackage{amsfonts}
\usepackage{amsmath}
\usepackage{amssymb}
\usepackage[sort&compress]{natbib}
\usepackage{dcolumn}
\usepackage{multirow}
\usepackage{bigstrut}
\usepackage{mathrsfs}
\usepackage{type1cm}
\usepackage{placeins}
\usepackage{color}
\usepackage{rotating}
\usepackage{slashed}
\usepackage{subfigure}
\begin{document}

\preprint{\today}

\title{Question of Lorentz invariance in muon decay}

\author{J. P. Noordmans}
\affiliation{Van Swinderen Institute for Particle Physics and Gravity,
                  University of Groningen,
                  9747 AG Groningen, The Netherlands}
\affiliation{CENTRA, Departamento de F\'isica, Universidade do Algarve, 8005-139 Faro, Portugal}
\author{C. J. G. Onderwater}
\author{H. W. Wilschut}
\author{R. G. E. Timmermans}
\affiliation{Van Swinderen Institute for Particle Physics and Gravity,
                  University of Groningen,
                  9747 AG Groningen, The Netherlands}

\date{\today}
\vspace{3em}

\begin{abstract}
\noindent
Possibilities to test the Lorentz invariance of the weak interaction in muon decay are considered.
We derive the direction-dependent muon-decay rate with a general Lorentz-violating addition to the $W$-boson propagator. 
We discuss measurements of the directional and boost dependence of the Michel parameters and of the muon lifetime as a function
of absolute velocity. The total muon-decay rate in the Lorentz-violating Standard Model Extension is addressed. Suggestions are
made for dedicated (re)analyses of the pertinent data and for future experiments. 
\end{abstract}

\maketitle

\section{Introduction}
Muon decay has historically been an important tool to establish the left-handed
``$V-A$'' Lorentz structure of the weak interaction in the development of the Standard Model (SM)
of particle physics. Nowadays, it is used to search for new interactions that arise in SM extensions
to energies above the electroweak scale~\cite{Kun01,Fet10,Gag05}. In recent years, the limits on
such contributions have been significantly improved~\cite{Mus05}. In this article, we add yet another
twist to muon decay: We propose muon decay as a precision laboratory to test the invariance of the
weak interaction under Lorentz transformations (boosts and rotations). Scenarios that break Lorentz
and CPT invariance occur in many proposals to unify the SM with general relativity~\cite{Lib13,Tas14},
a central open issue in high-energy physics.
While CPT invariance, in particular,  has been tested in neutrino and neutral-meson
oscillations, the available evidence for Lorentz invariance of weak decays is limited~\cite{Lee56}.
Further motivation for investigating the muon comes from experiments on the muon
anomalous magnetic moment (``$g-2$'')~\cite{Ben02} and muonic hydrogen~\cite{Poh10},
where at present puzzling deviations from the SM exist.

To explore Lorentz violation in weak decays and to guide
and interpret the pertinent experiments, an effective field theory approach was
developed~\cite{Noo13a,Noo13b,Vos15} in which Lorentz violation is parametrized by a complex tensor
$\chi^{\mu\nu}$ ($\chi^{\mu\nu}$ is CPT even or odd depending on its momentum
dependence~\cite{Noo13a}). This approach includes a wide class of Lorentz-violating effects, in particular
 contributions from a modified $W$-boson propagator
$\left\langle W^{\mu+}W^{\nu-}\right\rangle = -i(g^{\mu\nu}+\chi^{\mu\nu})/M_W^2$ or from a
Lorentz-violating vertex $-i\gamma_\nu(g^{\mu\nu}+\chi^{\mu\nu})$ \cite{note1}; $g^{\mu\nu}$ is the Minkowski
metric. Bounds on components of $\chi^{\mu\nu}$ were previously extracted from semileptonic
allowed~\cite{Wil13,Mul13} and forbidden nuclear $\beta$ decay~\cite{Noo13b} and pion
decay~\cite{Alt13,Noo14}, and from nonleptonic kaon decay~\cite{Vos14}. These bounds were translated
into limits on parameters of the Standard Model Extension (SME)~\cite{Col98,Kos11}, which is the most
general effective field theory for the breaking of Lorentz and CPT invariance. 

In this article we derive the muon-decay rate in our general framework. We show that muon decay offers many
possibilities to search for Lorentz violation and we discuss some general issues for dedicated laboratory
experiments. We give examples of Lorentz-violating observables and how they could be measured. From
a measurement by the TWIST Collaboration~\cite{Mus05} we extract bounds on components of $\chi^{\mu\nu}$.
From available data on the lifetime of muons at rest and in flight, we constrain the boost dependence of the muon
lifetime. We propose reanalyses of existing measurements and new muon-decay and muon $g-2$ experiments.
We argue that dedicated laboratory experiments with muons are preferred over observations of cosmic-ray muons.
Finally, we summarize our conclusions.

\section{Muon-decay rate}
When $\chi^{\mu\nu}$ is included in the $W$-boson propagator, the matrix element for the decay
$\mu^-\rightarrow e^-+\overline{\nu}_e+\nu_\mu$, corresponding to the tree-level $W$-exchange diagram,
reads ($\hbar=1=c)$
\begin{equation}
iM = \frac{G_F}{\sqrt{2}}(g^{\mu\nu}+\chi^{\mu\nu})
    \left[\bar{u}(k_1)\gamma_\mu(1-\gamma_5)u(l)\right]\left[\bar{u}(p)\gamma_\nu(1-\gamma_5)v(k_2)\right]\ ,
\label{matrixelmumin}
\end{equation}
where $G_F$ is the Fermi coupling constant, and $l$, $p$, $k_1$, and $k_2$
are the momenta of the muon, electron, muon neutrino, and electron antineutrino, respectively.
For simplicity, we only consider the dominant momentum-independent part of $\chi^{\mu\nu}$. Although this implies
$\chi^{\mu\nu} = \chi^{\nu\mu*}$ when $\chi^{\mu\nu}$ originates from the $W$-boson propagator, we keep contributions 
from the real-antisymmetric and the imaginary-symmetric parts of $\chi^{\mu\nu}$ for generality. Such contributions 
can result {\it e.g.} from a Lorentz-violating correction to the vertex \cite{note1}.

From the matrix element in Eq.~\eqref{matrixelmumin} and that for $\mu^+\rightarrow e^++\nu_e+\overline{\nu}_\mu$
we derive, to first order in Lorentz violation, the muon-decay rate
\begin{eqnarray}
dW &=& \frac{G_F^2}{24\pi^4}\frac{d^3 p}{2 l^0 2p^0}\Big[q^2 (L\cdot Q) + 2(L\cdot q)(Q\cdot q) \notag \\
 && + 2\chi_{rs}^{\mu\nu}\left(2q^2 L_\mu Q_\nu + (L\cdot Q)q_\mu q_\nu - (q\cdot Q)q_\mu L_\nu - (L\cdot q)Q_\mu q_\nu\right) \notag \\
 && + 2\chi_{ra}^{\mu\nu} \left(q^2 L_\mu Q_\nu - (q\cdot Q)q_\mu L_\nu - (L\cdot q)Q_\mu q_\nu\right) \notag \\
 && + \chi_{ia}^{\mu\nu}\epsilon_{\mu\nu\varrho\sigma}\left((q\cdot Q)L^\varrho q^\sigma -
 (L\cdot q)q^\varrho Q^\sigma\right) - 2\chi_{is}^{\mu\nu}q_\mu \epsilon_{\nu\varrho\sigma\lambda}L^\varrho q^\sigma Q^\lambda\Big]\ , 
\label{generaldecayrate}
\end{eqnarray}
where we summed over the spins and integrated over the momenta of the (anti)neutrino.
The subscripts $r$, $i$, and $s$, $a$ on the tensor $\chi^{\mu\nu}$ denote its real or imaginary
and its symmetric or antisymmetric part, respectively. We defined the four-vectors $q = l - p$,
$L = l \mp m_\mu s$, and $Q = p \mp m_e r$, where the upper (lower) sign applies for $\mu^-$
($\mu^+$) decay; $m_\mu$ and $m_e$ are the muon and electron mass, respectively. The spin
four-vector $s$ of the muon is
\begin{equation}
s = \left(\frac{{\bf l \cdot \hat{s}}}{m_\mu},{\bf \hat{s}}+\frac{({\bf l \cdot \hat{s}}){\bf l}}{m_\mu(l^0 + m_\mu)}\right)\ ,
\end{equation}
with ${\bf \hat{s}}$ a unit vector in the direction of the spin of the muon in its rest frame; the spin four-vector $r$
of the $\beta^\mp$ particle (electron/positron) is defined analogously.

When we sum Eq.~\eqref{generaldecayrate} over the spin of the $\beta$ particle
we obtain for the differential decay rate in the muon rest frame
\begin{eqnarray}
\frac{dW}{dx d\Omega} &=& \frac{W_0}{\pi}x^2\Bigg[3(1-x)+\frac{2}{3}\varrho(4x-3)
 \mp ({\bf\hat{s}}\cdot{\bf\hat{p}})\xi\left[1-x + \frac{2\delta}{3}(4x-3)\right] \notag \\ 
&& -\Big(t_1 + {\bf v}_1\cdot{\bf\hat{p}} \pm {\bf v}_2\cdot{\bf\hat{s}} \pm {\bf v}_3\cdot({\bf \hat{s}}\times{\bf \hat{p}}) \notag \\
&& + T_1^{ml}\hat{p}^m\hat{p}^l \pm T_2^{ml}\hat{p}^m\hat{s}^l \pm T_3^{ml}\hat{p}^m({\bf \hat{s}}\times{\bf \hat{p}})^l \Big)(1-x) \notag \\
&& -\Big(z_1 + {\bf u}_1\cdot{\bf\hat{p}} \pm {\bf u}_2\cdot{\bf\hat{s}} \pm {\bf u}_3\cdot({\bf \hat{s}}\times{\bf \hat{p}}) \notag \\
&& + H_1^{ml}\hat{p}^m\hat{p}^l \pm H_2^{ml}\hat{p}^m\hat{s}^l \pm H_3^{ml}\hat{p}^m({\bf \hat{s}}\times{\bf \hat{p}})^l\Big)(4x-3)\notag \\
&& \pm({\bf \hat{s}}\cdot{\bf \hat{p}})\left[\left(t_2+{\bf v}_4\cdot{\bf\hat{p}}\right)(1-x) + \left(z_2 + {\bf u}_4\cdot{\bf\hat{p}}\right)(4x-3)\right]
\Bigg]\ ,
\label{muondecayinx}%
\end{eqnarray}
where $W_0 = G_F^2 m_\mu^5/(192\pi^3)$ is the total SM decay rate, and $x = E/E_{\mathrm{max}}$ is the
energy of the $\beta$ particle relative to its maximum. We neglected terms proportional to $m_e/m_\mu$,
because the pertinent SM terms do not mimic Lorentz violation and Lorentz-violating terms proportional
to $m_e/m_\mu$ are suppressed. The Lorentz-violating parameters in Eq.~\eqref{muondecayinx}
are defined by
\begin{subequations}
\begin{eqnarray}
t_1 & = & z_1 = z_2 = \frac{1}{2}\chi_{rs}^{00}\ ,  t_2 = \frac{5}{2}\chi_{rs}^{00}\ ; \\
v_1^l & = & \chi_{rs}^{0l} + 2\chi_{ra}^{0l} - 2\tilde{\chi}_{ia}^l\ ,
v_2^l = \frac{1}{2}\chi_{rs}^{0l} + \frac{7}{2}\chi_{ra}^{0l} + \frac{5}{4}\tilde{\chi}_{ia}^{l}\ , \notag \\
v_3^l & = & \frac{3}{2}\chi_{ia}^{0l} + \frac{5}{2}\chi_{is}^{0l}\ ,
v_4^l = \frac{3}{2}\chi_{rs}^{0l} + \frac{3}{2}\chi_{ra}^{0l} - \frac{3}{4}\tilde{\chi}_{ia}^l\ ;  \label{VectorV} \\
u_1^l & = & -\frac{1}{2}\tilde{\chi}_{ia}^{l}\ , u_2^l = \frac{1}{2}\chi^{0l}_{rs} + \frac{1}{2}\chi_{ra}^{0l} + \frac{1}{4}\tilde{\chi}_{ia}^l\ , \notag \\
u_3^l & = & \frac{1}{2}\chi_{ia}^{0l} + \frac{1}{2}\chi_{is}^{0l} \ , u_4^l = \frac{1}{2}\chi_{rs}^{0l} + \frac{1}{2}\chi_{ra}^{0l} - \frac{1}{4}\tilde{\chi}^l_{ia}\ ; \\
T_1^{ml} & = & -\frac{3}{2}\chi_{rs}^{ml}\ , T_2^{ml} = \frac{7}{2}\chi_{rs}^{ml} + \frac{1}{2}\chi_{ra}^{ml}\ , T_3^{ml} = \frac{3}{2}\chi^{ml}_{is}\ ; \label{TensorT} \\
H_1^{ml} & = & -\frac{1}{2}\chi_{rs}^{ml}\ , H_2^{ml} = \frac{1}{2}\chi_{ra}^{ml}\ , H_3^{ml} = \frac{1}{2}\chi_{is}^{ml}\ , \label{TensorH}
\end{eqnarray}%
\label{LVquantitiesmuon}%
\end{subequations}
where $\tilde{\chi}^l=\epsilon^{lmk}\chi^{mk}$.
The first line of Eq.~\eqref{muondecayinx} gives the SM decay rate written in the conventional way~\cite{Fet10}.
For easy comparison, we inserted by hand the three standard Michel parameters $\varrho$, $\xi$, and $\delta$,
which parametrize the energy and angular distribution of the $\beta$ particles in polarized muon decay \cite{Mic50},
and which are used to test the $V-A$ structure of the weak interaction. In the SM [and in our framework
for Lorentz violation, {\em cf.} Eq.~\eqref{matrixelmumin}] the currents have $V-A$ structure, in which case the values
of the Michel parameters are $\varrho=3/4$, $\xi=1$, and $\delta=3/4$. The TWIST Collaboration has in recent years
put strong limits on non-($V-A$) contributions to the Michel parameters~\cite{Mus05}. The next lines
of Eq.~\eqref{muondecayinx}, with the parameters defined in Eqs.~\eqref{LVquantitiesmuon}, give Lorentz-violating,
frame-dependent contributions to muon decay.

\section{Bounds from the Michel parameters}
Equation~\eqref{muondecayinx} offers many possible tests of Lorentz invariance in muon decay. For example, the
dependence of the decay rate on the $\beta$ direction can be studied. In general, it is profitable to measure over extended
periods of time and record the data with ``time stamps.'' One can then search for signals that oscillate with periods of
one or one-half sidereal day due to the rotation of Earth with respect to the standard Sun-centered inertial reference
frame~\cite{Kos11,Noo13a}. This strategy requires reanalyses of, typically statistics-limited, existing data~\cite{Mus05}
or new dedicated experiments. Another option is to compare experiments performed at different velocities, {\it i.e.}
with different values for the Lorentz boost factor $\gamma$, because at higher $\gamma$ values the
Lorentz-violating signals are enhanced by a factor $\gamma^2$, so that with an equal number of events
more stringent limits can be set. We discuss three examples in more detail.

($i$) To illustrate the rotational dependence of muon decay, consider the decay rate of unpolarized muons depending
on the direction of the outgoing $\beta$ particles. If one measures the number of events in two detector halves, each
spanning $2\pi$ of solid angle, one can determine an asymmetry, to first order in Lorentz violation, given by
\begin{equation}
\mathcal{A} = \frac{N_+ - N_-}{N_+ + N_-} = -\frac{1}{6}{\bf v}_1\cdot{\bf \hat{n}}\ ,
\label{asymmetry}
\end{equation}
where $N_{\pm}$ is the number of particles emitted in the hemisphere with its axis in the $\pm {\bf \hat{n}}$ direction, while the
``preferred direction'' ${\bf v}_1$ in the muon rest frame is defined in Eq.~\eqref{VectorV}.
When the laboratory $\hat{z}$ axis is perpendicular to Earth's surface, $\hat{x}$
points south, and $\hat{y}$ points east, the relation between ${\bf v}_1$ in the laboratory frame and ${\bf V}_1$
in the Sun-centered frame is given by~\cite{Noo13a}
\begin{equation}
{\bf v_1} = \left(\begin{array}{c}v_1^x \\ v_1^y \\ v_1^z\end{array}\right) = \left(\begin{array}{c}\cos\zeta \cos\Omega t\;V^x_1
+ \cos\zeta \sin \Omega t\; V^y_1 - \sin\zeta\; V^z_1 \\ -\sin\Omega t\; V^x_1 + \cos\Omega t\; V^y_1 \\ 
\sin\zeta \cos\Omega t\; V^x_1 + \sin\zeta \sin\Omega t\; V^y_1 + \cos\zeta\; V^z_1 \end{array}\right)\ ,
\label{frametranslation}
\end{equation} 
where $\zeta$ is the colatitude of the site of the experiment on Earth and $\Omega\simeq 2\pi/$(23h56m) is the angular
rotation frequency of Earth.
\begin{figure}[t]
\centering
\includegraphics[width=0.60\textwidth]{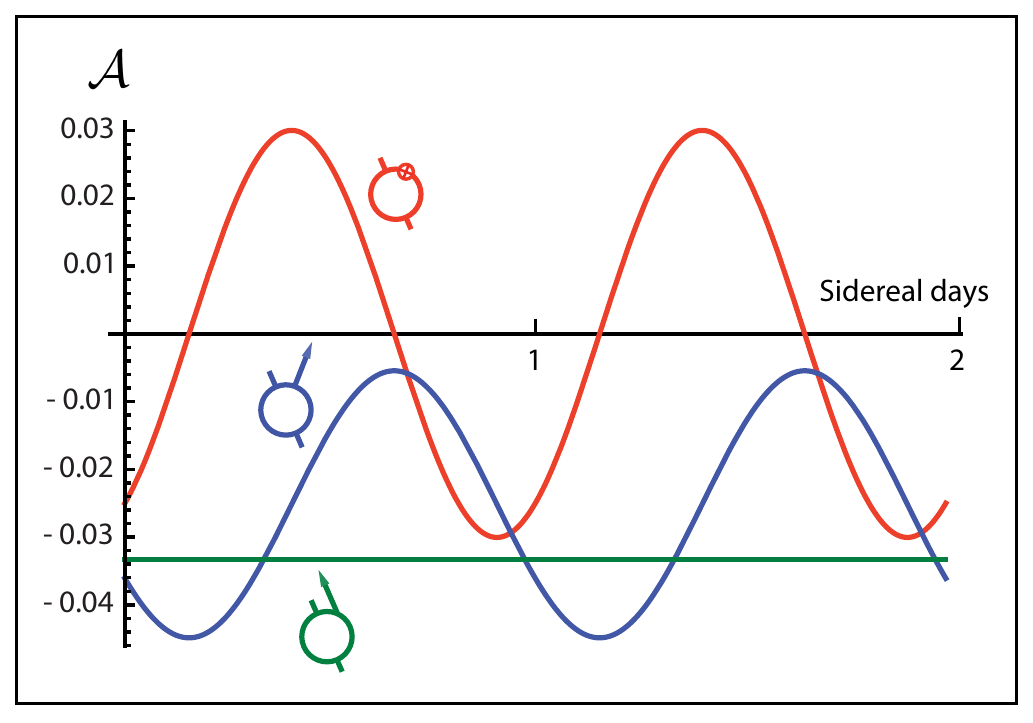}
\caption{The asymmetry $\mathcal{A}$ of Eq.~\eqref{asymmetry} as a function of sidereal time, for $V_1^x = 0.1$, $V_1^y = 0.15$,
                $V_1^z = 0.2$, and $\zeta = 41^{\circ}$. Its amplitude and offset are determined by the direction of the $\beta$ particle,
                as depicted relative to Earth's axis. Present limits indicate ${\mathcal{A}} < \mathcal{O}(10^{-4}$).}
\label{fig:oscsign}
\end{figure}
Equation~\eqref{frametranslation} shows that observables like $\mathcal{A}$ will oscillate with a period of one sidereal day.
In Fig.~\ref{fig:oscsign} three possible scenarios for ${\bf v}_1\cdot{\bf \hat{n}}$ are illustrated. The green line shows no
oscillation, since the axis of the detector halves is parallel to Earth's axis. The red line is for the case where this axis points
east, {\it i.e.} perpendicular to Earth's axis, in which case the offset of the oscillation is zero. This can help to identify
systematic effects, since any constant offset is not due to the ${\bf v}_1\cdot{\bf \hat{n}}$ term. For the blue line, where
${\bf \hat{n}}$ is perpendicular to Earth's surface, there is an oscillation as well as an offset.

($ii$) When we integrate Eq.~\eqref{muondecayinx} over the energy of the $\beta$ particle, all terms proportional to
$4x-3$ disappear. Defining the muon-polarization direction as the $z$ axis and integrating over the azimuthal angle
$\phi$ of the $\beta$ momentum, we find
\begin{eqnarray}
\frac{dW}{d\cos\theta} & = & \frac{W_0}{6}\Big[3 - t_1 \mp v_2^z \mp \cos\theta\left(\xi - t_2 \pm v_1^z + T_2^{zz} \right) \notag \\ 
&& - \cos^2\theta \left(T_1^{zz} \mp v_4^z \right) -\frac{1}{2}\sin^2\theta \left(T_1^{xx}+T_1^{yy}\right)\Big]\ ,
\label{muonrateobservables}
\end{eqnarray}
where $\theta$ is the angle between the polarization axis and the $\beta$ momentum, with the muon at rest
in the laboratory frame. Thus, when one determines the Michel parameter $\xi$ by fitting the $\theta$
dependence of the decay rate, a term with $\cos^2 \theta$ dependence has to be included. Since
the Lorentz-violating coefficients of the $\theta$-dependent terms vary over the course of a sidereal day,
one has to express them in terms of $X^{\mu\nu}$, by which we denote $\chi^{\mu\nu}$ in the Sun-centered
frame, as in Eq.~\eqref{frametranslation}, and integrate over the relevant measurement periods. Some observables
that depend on parameters with two spacelike indices, {\it i.e.} $T_i^{ml}$ and $H_i^{ml}$ in Eqs.~\eqref{TensorT}
and~\eqref{TensorH}, will oscillate in addition with a period of half a sidereal day.

($iii$) The TWIST value of the Michel parameter $\varrho_{\rm exp} = 0.74977(26)$ for $\mu^+$ decay~\cite{Mus05}
can already be used to derive a bound on $\chi^{\mu\nu}$.  The decay rate as a function of positron energy, without selecting
a particular direction for the positrons, follows from Eq.~\eqref{muondecayinx} as
\begin{equation}
\frac{dW}{dx} = 4W_0x^2\left[3(1-x)(1+ {\bf n}_1\cdot{\bf \hat{s}})
+\left(\frac{2}{3}\varrho-\frac{1}{3}\chi_{rs}^{00}+{\bf n}_2\cdot{\bf \hat{s}}\right)(4x-3)\right]\ ,
\label{decaywithx}
\end{equation}
with $n_1^l=\chi^{0l}_{ra} + \tfrac{1}{2}\tilde{\chi}_{ia}^{l}$ and $n_2^l =
                         \tfrac{1}{3}(\chi_{rs}^{0l}+\chi_{ra}^{0l}+\tilde{\chi}_{ia}^l)$.
When the difference between the measured value and the SM prediction is attributed to $\chi^{\mu\nu}$, we get
$\varrho_{\rm exp} = \varrho_{\rm SM} - \chi_{rs}^{00}/2+(3{\bf n}_2/2-\varrho_{\rm SM}\,{\bf n}_1)\cdot{\bf\hat{s}}$,
which results in the 95\% confidence limit (C.L.)
\begin{equation}
-5.6 \times 10^{-4} < X_{rs}^{00}+\sin\zeta\cos\phi
   \left(X_{rs}^{0Z}-\frac{1}{2}X_{ra}^{0Z}+\frac{1}{4}\tilde{X}_{ia}^Z\right) < 1.5 \times 10^{-3}\ ,
\label{limitrho}
\end{equation}
where $\zeta \simeq 41^\circ$ is the colatitude of Vancouver, and $\phi\simeq 52^\circ$ is the angle that the
$\mu^+$ momentum, when entering the TWIST solenoid, makes with the north-south direction (anticlockwise) in
the plane parallel to the surface of Earth, so $\sin\zeta\cos\phi=0.40$ (the $\mu^+$ spin points opposite to its
momentum). Eq.~\eqref{limitrho} indicates that with dedicated analyses, along the lines of examples ($i$) and
($ii$), and realistic statistics muon decay can improve on the existing bounds of order
$\leq\mathcal{O}(10^{-4}$)~\cite{Kos11}.

\section{Boost dependence of the muon-decay rate}
Strong bounds on Lorentz violation can be obtained by comparing muon-lifetime measurements
at different absolute muon velocities, {\it i.e.} different Lorentz boost factors $\gamma$. The data
that are available, unfortunately, are only for the total muon decay rate $W$, {\it i.e.} the muon lifetime.
For a matrix element of the form of Eq.~\eqref{matrixelmumin}, two pure $V-A$ currents contracted
by $g^{\mu\nu} + \chi^{\mu\nu}$, a general argument~\cite{Nie82,NooPhD}, summarized in the Appendix,
shows that the total decay
rate of unpolarized muons depends neither on $\chi_s^{\mu\nu}$ nor on $\chi_a^{\mu\nu}$, while the
total decay rate of polarized muons depends only on $\chi_a^{\mu\nu}$. These conclusions are borne
out by an explicit calculation starting from Eq.~\eqref{muondecayinx}. A residual dependence on the muon
polarization or on the electron/positron direction, however, could introduce a dependence on $\chi^{\mu\nu}$
(see next section).

\begin{table}[t]
\centering
\setlength{\tabcolsep}{10pt}
\begin{tabular}{c|cc|cc|c}
\hline\hline
&  $\tau$ ($\mu$s) & Ref. & $\tau'/\gamma$ ($\mu$s) & Ref. & $10^4\,\Delta$  \\
\hline
  $\mu^+$ & $2.1969803(22)_{\rm tot}$ &  \cite{Chi07} &
                      $2.1966(20)_{\rm tot}$      & \cite{Bai79}  & $-1.7(9.1)_{\rm tot}$ \\
  $\mu^-$ & $2.196998(31)_{\rm tot}$  & \cite{And13}  &
                     $2.1948(10)_{\rm tot}$    &  \cite{Bai79}  & $-10.0(4.6)_{\rm tot}$  \\
  $\mu$    & $2.1969804(22)_{\rm tot}$  &                         &
                     $2.19516(89)_{\rm tot}$       &                         & $-8.3(4.1)_{\rm tot}$ \\
\hline\hline
\end{tabular}
\caption{Muon lifetimes in $\mu$s at rest, $\tau$, and in flight at $\gamma\simeq29.3$, $\tau'$, and
the relative difference $\Delta=(\tau'/\gamma-\tau)/\tau$. Listed in the rows are the values for $\mu^+$,
$\mu^-$, and their average ($\mu$). The numbers between parentheses in the entries are the total errors.}
\label{tab:lifetime}
\end{table}

The available data for the $\mu^\pm$ lifetime that we used are collected in Table~\ref{tab:lifetime}. The most
precise measurement of the $\mu^+$ lifetime at rest, $\tau=1/W$, comes from the MuLan experiment
\cite{Chi07}. The $\mu^-$ lifetime at rest was derived from the MuCap experiment~\cite{And13}
by correcting for the muon capture rate on hydrogen, for which we used the theoretical value
$\Gamma(\mu^-p\rightarrow\nu_\mu n)=718(7)$ s$^{-1}$~\cite{Pas13}, obtained in chiral perturbation
theory, the low-energy effective field theory of QCD. We also list in Table~\ref{tab:lifetime} the averaged
$\mu^+$ and $\mu^-$ lifetimes as $\mu$, which is relevant when assuming CPT invariance. The muon
lifetime in flight, $\tau'$, was obtained from data published by the CERN $g-2$ Collaboration \cite{Bai79,Qia06}.
In this experiment, which was performed at the muon ``magic momentum,'' corresponding to $\gamma \simeq 29.3$,
the muons are kept in a circular orbit (the effects of acceleration on the lifetime are claimed to be negligible,
{\it cf.} Refs.~\cite{Bai77,Eis87}). The arrival times and energies of the $\beta$ particles were recorded together
with the magnetic-field strength. From these, the dilated lifetime $\tau'$ and $\gamma = 29.327(4)$ were
obtained \cite{Bai79}.

The error in $\tau'$ is dominated by statistics. The main systematic error is due to unknown gain variations
in the electron detectors, which result in time-dependent variations in the detection efficiencies. Muon losses,
caused {\it e.g.} by muon scraping on the ring, and a background of stored protons contribute significantly less
(for $\mu^-$, the effect of stored antiprotons was negligible). The average of the $\mu^+$ and $\mu^-$ lifetimes
and its errors is calculated by weighing with the inverse of the square of the error.

To determine bounds on Lorentz violation, we compare the lifetime for muons in flight to the one obtained
for muons at rest. When Lorentz invariance holds, the muon lifetime at rest is calculated from $\tau = \tau'/\gamma$,
therefore $\Delta = (\tau'/\gamma-\tau)/\tau$ is the relevant dimensionless quantity. By using the values for $\tau'$
and $\gamma$ of the $g-2$ experiment, as given in Ref.~\cite{Bai79}, together with the values for $\tau$ in
Table~\ref{tab:lifetime}, we calculated the values for $\Delta$ for $\mu^+$, $\mu^-$, and for the average of $\mu^+$
and $\mu^-$, respectively, as listed in the last column of Table~\ref{tab:lifetime}. All results are consistent with zero,
although there is some mild stress for the negative muon, where $\Delta$ deviates from zero by 2.2$\,\sigma$. To
test CPT invariance of the $\mu^+$ and $\mu^-$ lifetimes, we consider the ratio
$R=2(\tau_{\mu^+}-\tau_{\mu^-})/(\tau_{\mu^+}+\tau_{\mu^-})$. For muon decay at rest and in flight we find
$R=-0.8(1.4)\times 10^{-5}$ and $R=8.2(10)\times 10^{-4}$, respectively.

\section{Interpretation}
The measurements and analyses of the
muon lifetime at rest and in flight, discussed in the previous section, were designed to be sensitive only to the
total unpolarized muon decay rate. In order to properly investigate the presence of Lorentz violation according to
Eq.~\eqref{muondecayinx}, details of the analyses from which the total lifetimes were extracted are required, since
these analyses involve taking averages over the muon direction and spin. For instance, it becomes relevant that
the muons in some measurements may have been polarized, even on average. This could have been due
to, for instance, a small residual polarization of the muons used in the analyzed MuLan and muon $g-2$ data sets, or
to a polarization component along the magnetic field in the muon $g-2$ data sets, which does not average out.

Such effects can be important, in particular, for the muon lifetime in flight, since the related Lorentz-violating contributions
are enhanced by boost factors. In the muon $g-2$ experiments, the arrival-time distribution of the $\beta$ particles is modulated
due to the muon-spin precession relative to its momentum. The analysis as reported in Ref.~\cite{Bai79} was such that the
result for $\tau'$ is predominantly sensitive to the exponential decay rather than this modulation. Furthermore, the fit to the
exponent of the decay curve is sensitive to the decay rate, independent of the direction or energy of the outgoing $\beta$
particles, even though the detectors are only sensitive to part of this parameter space. Also most effects of the muon polarization
are removed, because it precesses around the magnetic field and the muons are unpolarized on average, hence the net
transverse component of the polarization vanishes. However, any polarization component parallel to the magnetic field
does not average out and may thus persist as a residual vertical polarization. When averaged over a sidereal day, a possible
effect due to this residual polarization is further reduced as only the component along Earth's axis remains.
Taking such an effect into account, the decay rate is given by
\begin{equation}
W = \frac{1}{\gamma}W_0(1 \mp \gamma \cos\zeta\, \mathcal{P}_{\parallel} N_1^Z)\ ,
\label{residualpolar}
\end{equation}
where $\mathcal{P}_{\parallel}$ is the residual polarization parallel to the magnetic field oriented vertically, $\zeta$ is the
colatitude of the experiment, and ${\bf N}_1$ is ${\bf n}_1$ defined below Eq.~\eqref{decaywithx} in the Sun-centered frame.
Because the velocity of the muon is perpendicular to the magnetic-field directions, only one factor of $\gamma$ appears
in Eq.~\eqref{residualpolar}. Effects of incomplete rotation cycles of the muon, the muon spin, and incomplete sidereal
days are estimated 
to be suppressed by several orders of magnitude relative to the Lorentz-violating effect in Eq.~\eqref{residualpolar}.

From the available information we cannot assess whether such residual sensitivities would result in limits for the components
of $\chi_{\mu\nu}$ that can compete with the existing bounds. Therefore, our findings should
motivate more complete reanalyses of the existing data that differentiate the directions of the $\beta$ particles
and consider in detail residual polarizations of the muons. Moreover, dedicated new experiments are called for.

\section{Total muon-decay rate in the SME}
Since $\chi^{\mu\nu}$ does not contribute to the total decay rate, we briefly consider two  tensors from the minimal Standard
Model Extension (mSME), {\it i.e.} the power-counting renormalizable part of the full SME~\cite{NB}. The first one is the tensor
$c^{\mu\nu}$ that originates from the CPT-even part of the mSME Lagrangian, {\it viz.}
\begin{equation}
\mathcal{L} = c_{\mu\nu}\left[i\bar{\ell}\gamma^\mu \partial^\nu \ell
+ i\bar{\nu}\gamma^\mu \partial^\nu \nu + \bar{\ell}_L W^{\nu+}\gamma^\mu \nu_{L}
+ \bar{\nu}_L W^{\nu-}\gamma^\mu \ell_{L}\right]\ ,
\label{LVlagrleptoncmunu}
\end{equation}
where $\ell$ is the charged-lepton field, in our case the muon, and $\nu$ is the corresponding neutrino. The second one
is the vector $b^\mu$ that comes from a CPT-odd Lorentz-violating interaction involving the vectors $a_L^\mu$ and
$a_R^{\mu}$. Taking the flavor-diagonal part of the SME Lagrangian~\cite{Col98} and redefining the muon and muon-neutrino
field by $\psi \rightarrow \left[1-\tfrac{i}{2}(a^\mu_L+a^\mu_R)x_\mu\right]\psi$, the relevant part of the Lagrangian reads,
to first order in Lorentz violation,
\begin{equation}
\mathcal{L} = \bar{\ell}(i\slashed{\partial}-m-\gamma_5\slashed{b})\ell +
\bar{\nu}_L(i\slashed{\partial}-\slashed{b})\nu_L
+ \bar{\ell}_L\slashed{W}^+\nu_L + \bar{\nu}_L \slashed{W}^-\ell_L\ ,
\label{LVbyb}
\end{equation}
where $b^\mu = \tfrac{1}{2}(a_L^\mu - a_R^\mu)$ multiplies a dimension-three operator. (Since in the absence of interaction terms
with the $W$ boson, the neutrino field could be redefined such that the Lorentz-violating parameter would disappear from the
neutrino part of the Lagrangian, it is unobservable when only neutrinos are detected. The vector $b^\mu$ therefore cannot be
constrained by neutrino oscillations or time-of-flight measurements on neutrinos. In Ref.~\cite{Dia13} this point was
discussed for the equivalent electron-neutrino parameters.)

Since Eqs.~\eqref{LVlagrleptoncmunu} and \eqref{LVbyb} consist of free-fermion operators for external particles, some care has
to be taken in the calculation of the muon-decay rate. Following the procedures developed in Refs.~\cite{Col01,NooPhD}, one finds
that the total muon-decay rate is given by $W =W_0\left(1-21c_{00}/5\right)$ and $W = W_0\left(1\mp 4b^0/m_\mu\right)$ for $c^{\mu\nu}$
and $b^\mu$, respectively. Since we work to first order in Lorentz violation the effects of $c^{\mu\nu}$ and $b^\mu$ can be added
perturbatively. When boosted to a frame wherein the muons are moving, the resulting decay rate becomes
\begin{equation}
W = \frac{1}{\gamma}W_0\left[1-\frac{21 \gamma^2}{5}\left(\frac{c^{\mu\nu}p_\mu p_\nu}{(p^0)^2}\right) \mp
        \frac{4\gamma b^\alpha p_\alpha}{p^0 m_\mu}\right] \ , \\
\end{equation}
where $c^{\mu\nu}$ and $b^\mu$ are defined in the laboratory frame and $p$ is the muon momentum in this frame. Averaging this
over a rotation of the muon around the ring and over a full sidereal day we find in the Sun-centered frame
\begin{equation}
\int^{2\pi}_{0}\frac{d\phi}{2\pi}\int^{2\pi/\Omega}_{0}\frac{\Omega dt}{2\pi}W'
\simeq \frac{1}{\gamma}W_0\left[1-5.8\gamma^2(c^{TT}-0.1 c^{ZZ})\mp \frac{4\gamma b^T}{m_\mu}\right]\ ,
\label{finalinflightforb}
\end{equation}%
where $\Omega$ is the angular rotation frequency of Earth, and we used that the colatitude of CERN is about $43.8^\circ$
and that $\gamma^2 \gg 1$. We neglected effects of incomplete rotation cycles of the muon, the muon spin, and effects of
incomplete sidereal days. These effects are estimated to be 
several orders of magnitude smaller than the Lorentz-violating effect in Eq.~\eqref{finalinflightforb}. 
\begin{figure}[t]
\includegraphics[width=0.65\textwidth]{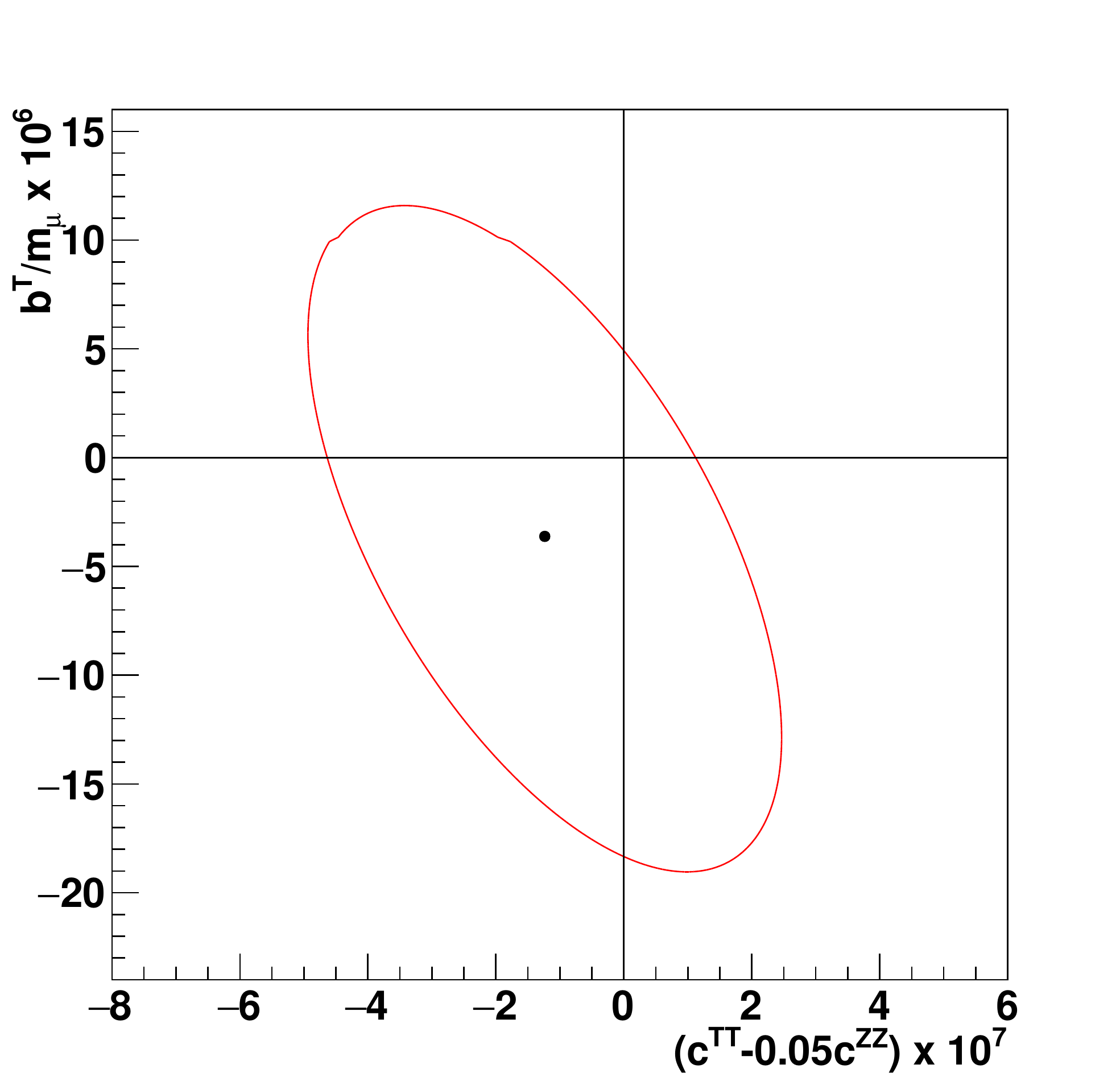}
\caption{Limits on CPT-even and CPT-odd Lorentz-violating couplings. The ellipse
                shows the joint 95\% C.L. region. The most likely point corresponds to Eq.~\eqref{cbvalues}.}
\label{CPTevenCPTodd}
\end{figure}
%

In Fig.~\ref{CPTevenCPTodd} we plot the joint $95\%$ C.L. region of $c^{TT}-0.1 c^{ZZ}$ and $b^T$ for a fit to the data
points of Table~\ref{tab:lifetime}. The most likely point in this parameter space is given by
\begin{eqnarray}
c^{TT}-0.1 c^{ZZ} & = & -1.2(1.1)\times 10^{-7}\ , \notag \\
b^T/m_\mu & = & -3.6(4.5)\times 10^{-6}\ ,
\label{cbvalues}
\end{eqnarray}
with $1\,\sigma$ errors. It is represented by the dot at the center of the ellipse in Fig.~\ref{CPTevenCPTodd}.
To our knowledge, these are the first values for the $b^T$ coefficient of the second-generation leptons. 
The space components of $b^\mu$ have been bounded to a level of $10^{-23}$-$10^{-24}$ GeV by analyzing
the spin-precession frequency of muons~\cite{Ben08}. The values in Eq.~\eqref{cbvalues} are an order
of magnitude larger than the bounds on the neutrino coefficients derived in Ref.~\cite{Dia13}, but these bounds
are for first-generation coefficients.

\section{Cosmic-ray muons}
Because large $\gamma$ factors are advantageous, cosmic-ray muons, which can have
energies up to at least $10^4$ GeV, are obvious candidates to search for Lorentz violation.
In Ref.~\cite{Col97}, it was pointed out that the rate of the flavor-violating muon-decay
mode $\mu \rightarrow e + \gamma$ could be enormously enhanced when Lorentz
invariance is violated. Strong bounds on Lorentz violation for this decay mode
were subsequently obtained in Ref.~\cite{Cow99}. Therefore, it is of interest to discuss
here whether cosmic-ray muons could also be used to put strong limits on Lorentz violation
for the ordinary weak decay of the muon.

When dealing with such ultrahigh energies, one has to address the question up to
which energy the theoretical framework is valid. Frames that move relatively slow
with respect to Earth are called concordant frames~\cite{Kos00}. In these frames
all Lorentz-violating parameters are expected to be small. However, in frames that
are highly boosted with respect to concordant frames, the Lorentz-violating parameters
can become so large that they cause problems with stability and causality in the
theory. For very large boost factors the muon-decay rate could even
become negative. When we denote the dimensionless Lorentz-violating effect in
the muon rest frame by $A$, then for large $\gamma$ the tensor $\chi^{\mu\nu}$
with two Lorentz indices scales schematically as $A\propto \gamma^2 a$, where $a$
is the Lorentz-violating effect in a concordant frame. A large $\gamma$ factor gives
better bounds on $a$ when we have a bound for $A$.  However, the theory can only
be trusted up to some $A = A_{\mathrm{max}}$. If we take as a guideline
$A_{\mathrm{max}} = 10^{-2}$~\cite{Kos00}, the decay rate has to be determined with
subpercent precision to get reliable bounds. This kind of precision is very hard to achieve
for ultrahigh-energy cosmic-ray muons. Therefore, results for the boost dependence of the decay
rate of such high-energy muons are hard to relate to Lorentz-violating coefficients in an effective
field theory approach.

This is less of a problem for the analysis in Ref.~\cite{Cow99}, because $\mu \rightarrow e + \nu + \nu$
is the main allowed decay mode in the SM, while $\mu \rightarrow e + \gamma$ is a forbidden process
that gets enhanced with respect to the SM decay mode, even without a large boost factor.  Moreover, the
amplitude for $\mu \rightarrow e + \gamma$ does not interfere with a SM amplitude,
because it is not a correction to an already existing SM process. Although it therefore depends quadratically
on a Lorentz-violating parameter, the enhancement with $\gamma$ will also be squared, resulting in a scaling
with $\gamma^4$. For $\mu \rightarrow e + \gamma$ Lorentz violation could thus become detectable for
values of the boost factor $\gamma$ for which the Lorentz-violating effect is still sufficiently small.

\section{Conclusion}
We derived the most general Lorentz-violating muon-decay rate in the context of our theoretical
framework~\cite{Noo13a,Noo13b}. Our main result, given by Eqs.~(\ref{muondecayinx}) and
(\ref{LVquantitiesmuon}), offers a wealth of possible precision tests of Lorentz invariance of the weak
interaction in muon decay. From the measurement of the Michel parameter $\varrho$ we derived
bounds of order $10^{-3}$ on $X_{\mathrm{rs}}^{00}$. Similar bounds on $X^{\mu\nu}$ could be
obtained  from the measurements of other Michel parameters. However, this requires reanalyses of
existing data~\cite{Mus05} or dedicated new measurements of the Michel parameters with higher
statistics. We gave examples of the types of analyses that are required for Lorentz-violating observables.
We compared the lifetime of muons at rest to the lifetime derived from measurements on muons in flight,
and derived bounds of order $10^{-6}$-$10^{-7}$ on specific parameters in the mSME.

The available data from refereed publications are consistent with Lorentz invariance. We advocated
reanalyses of the data in terms of $X^{\mu\nu}$ that take into account the dependence of the muon-decay
rate on the directions of the $\beta$ particles and the polarization of the muons. Analyses in terms of
nonmininal SME operators, along the lines of Refs.~\cite{Kos13,Gom14}, would then also be of interest.
Ultrahigh-energy cosmic-ray muons are not a good substitute for the experiments that we suggest, since
the decay rates would have to be measured with subpercent precision to get reliable bounds on Lorentz
violation. We suggest a dedicated analysis of the muon lifetime in the planned new muon $g-2$
experiment~\cite{Win13}. Based on the ${\cal O}(10^{12})$ expected muons, an order
of magnitude improvement on the dilated lifetime appears feasible. Since, apart from the required precision,
the main challenge in future experiments is to obtain competitive statistics, the muon-beam facilities under
consideration for a new generation of collider experiments are most relevant \cite{Pea05}.

\section*{Acknowledgments}
We thank A. Olin for helpful discussions on the details of the TWIST experiments and P. Cushman for
constructive discussions about the data of Ref.~\cite{Qia06}.
This research was supported by the Dutch Stichting voor Fundamenteel Onderzoek der Materie (FOM)
under Programmes 104 and 114 and project 08PR2636.
J. N. acknowledges the financial support of the Portuguese Foundation for Science and Technology (FCT)
under grant SFRH/BPD/101403/2014 and program POPH/FSE.  

\section*{Appendix: Total muon-decay rate}
Consider the square of the matrix element
$\left[\bar{u}_a\mathcal{O}_\mu u_b\right]\left[\bar{u}_c\mathcal{O}_\nu u_d\right]$, where
$\mathcal{O}_\mu = \gamma_\mu(1-\gamma_5)$ is a pure $V-A$ current. After Fierz transforming and contracting
with $g^{\mu\nu}$, $\chi_s^{\mu\nu}$, and $\chi_a^{\mu\nu}$, it follows that if one does not observe the momenta
and spins of particles $a$ and $c$, the symmetric part $\chi_s^{\mu\nu}$ does not enter the expression for the
decay rate. If, in addition, the momentum and spin of particle $d$ and the spin of particle $b$ are unobserved,
the terms with the antisymmetric part $\chi_a^{\mu\nu}$ in the decay rate also vanish. Applying this to
muon decay, it means that the total unpolarized decay rate cannot depend on $\chi^{\mu\nu}$, since it
cannot depend on $\chi_s^{\mu\nu}$, nor on $\chi_a^{\mu\nu}$. It also implies that the total decay rate
of polarized muons can only depend on $\chi_a^{\mu\nu}$.


\begin{thebibliography}{99}
\bibitem{note1} When $\chi^{\mu\nu}$ is taken as a vertex correction, gauge invariance implies
Lorentz-violating corrections for the corresponding external legs. These are not included in
Eq.~\eqref{generaldecayrate}.
\bibitem{NB} The operators considered in this section affect the properties (dispersion
relation, spinors, phase space) of the external particles. In particular, they influence the
cyclotron and the spin-precession frequency of the muon and the ratio of the muon and
proton magnetic moments as measured in muonium, which all enter the determination of
$\gamma$~\cite{Bai79}. However, these effects can be neglected here, because they are
not enhanced by a factor $\gamma$ or $\gamma^2$.
\end{thebibliography}
\end{document}